# The study of nearest- and next-nearest-neighbor magnetic interactions in seven tetragonal compounds of V(IV) containing linear chains and square lattices

**L M Volkova and S A Polyshchuk**

Institute of Chemistry, Far Eastern Branch of the Russian Academy of Sciences 690022 Vladivostok, Russia

E-mail: volkova@ich.dvo.ru

**Abstract**
A new crystal chemical method was used to calculate the sign and strength not only of the nearest-neighbor (NN) interactions, but also of the next-nearest-neighbor (NNN) ones in tetragonal compounds $Zn_2(VO)(PO_4)_2$ (I), $(VO)(H_2PO_4)_2$ (II), $(VO)SiP_2O_8$ (III), $(VO)SO_4$ (IV), $(VO)MoO_4$ (V), $Li_2(VO)SiO_4$ (VI) and $Li_2(VO)GeO_4$ (VII) with similar sublattices of $V^{4+}$ ions on the basis of the room-temperature structural data. The reason for difference between respective magnetic interactions characteristics of these compounds was established. It is shown that the characteristic feature of these compounds is a strong dependence of the strength of magnetic interactions and the magnetic moments ordering type on slight displacements of $XO_4$ (X = P, Mo, Si or Ge) groups even without change of the crystal symmetry. In addition to extensively studied square lattice, other specific geometrical configurations of $V^{4+}$ were discovered. These configurations can result in frustration of magnetic interactions, namely linear chains along the *c*-axis with competing nearest- and next-to-nearest-neighbor interactions; rectangular (in I) and triangular (in II–VII) lattices with non-equivalent nearest-to-neighbor interactions, which can be also considered as n-leg ladders; one extra square lattice in the *ab*-plane with longer range interactions. It was concluded that virtually all magnetic interactions in these compounds were frustrated.

## 1. Introduction

The search for and study of frustrated magnets are of great interest in determination of the role of frustrating interactions in formation of the magnetic state of low-dimensional systems. Recently, considerable attention has been paid to $(VO)MoO_4$, $Li_2(VO)SiO_4$ and $Li_2(VO)GeO_4$ [1–7] as prototypes of a frustrated two-dimensional S=1/2 antiferromagnet on a square lattice with competing interactions along the side and diagonal of the square. However, interpretations of experimental results and theoretical models of different researchers are in rather poor agreement with each other. Besides, although the existence of a linear VO-VO chain along the *c* axis assumes the presence of strong antiferromagnetic interactions in this direction, only in earlier works [8, 9] was one of these compounds (VO)MoO$_4$ considered as a one-dimensional antiferromagnet. It was shown later [1] that this system is essentially two-dimensional, with extremely weak interplanar coupling. The absence of strong AF interactions in linear chains VO-VO can be explained only by competition with other interactions. Hence, in order to reveal the magnetic state of (VO)MoO$_4$, Li$_2$(VO)SiO$_4$ and Li$_2$(VO)GeO4, it is necessary to consider frustration of magnetic interactions not only on a square lattice selected in [1-7], but also on other geometrical configurations of V$^{4+}$ ions existing in these compounds.

In this work, we have studied seven tetragonal compounds: Zn$_2$(VO)(PO$_4$)$_2$ [10], (VO)(H$_2$PO$_4$)$_2$ [11], (VO)SiP$_2$O$_8$ [12], (VO)SO$_4$ [13], (VO)MoO$_4$ [14], Li$_2$(VO)SiO$_4$ [15], and Li$_2$(VO)GeO$_4$ [15], with similar sublattices of V$^{4+}$ ions. The magnetic properties of the first four compounds have not been studied in detail. For each of these compounds we calculate the sign and strength of magnetic interactions not only between nearest neighbors, but also for longer-range neighbors by a new crystal chemical method [16] on the basis of structural data. The characteristics of magnetic interactions found can become apparent only in cases when there are no obstacles to their simultaneous existence resulting from geometrical configurations in the magnetic ion sublattice. We have analyzed the structure of V$^{4+}$ ion sublattice and selected such specific configurations that can cause frustration of magnetic interactions in compounds under study.

## 2. Method

The sign and strength of magnetic interactions in compounds under study were calculated by a new crystal chemical method [16] on the basis of structural data with using the program "MagInter". We have developed this method to estimate characteristics of magnetic interactions between magnetic ions located at any distances from each other. The method is phenomenological, since it does not take into account overlap of electronic shells of interacting ions, and it is based on known regularities. According to these regularities, the physical properties as well as the crystal structure of compound are determined by its electronic structure. Consequently, the compound's crystal structure can be applied to determine its physical properties. From the geometrical point of view, the Goodenough method [17] is a particular case of this method, when the length of bridging distances in a fragment $M$–$X$–$M$ is close to the length of covalent bonds $M$–$X$.

In [16] we have shown that the strength of magnetic interactions and the ordering type of the magnetic moments in low-dimensional crystal compounds are determined mainly by the geometrical arrangement and the size of the intermediate ion $A_n$ in the bounded space region between two magnetic ions $M_i$ and M$_j$. The bounded space region between $M_i$ and $M_j$ ions along the line of their interaction is defined as a cylinder whose radius is equal to that of these magnetic ions. Here we take into account not only anions, which are valent bound with the magnetic ions, but also all the intermediate negatively or positively ionized atoms, except cations of metals without unpaired electrons. The distances between magnetic ions, such as inside the low-dimensional fragment and between fragments, have an influence on the strength of magnetic interactions, however, they do not play a determining role in the interaction formation in case of absence of a direct interaction contribution.

If some intermediate ions enter into space between two magnetic ions, each of them, depending on the location, tends to orient the magnetic moments of these ions accordingly and makes a contribution to occurrence of antiferromagnetic or ferromagnetic components of magnetic interaction. The sign and value of the strength of interaction $J_{ij}^s$ between magnetic ions $M_i$ and $M_j$ is determined by the sum of these contributions $j_n^s$:

$$J_{ij}^s = \sum_n j_n^s .  \qquad (1)$$

If $J_{ij}^s < 0$, the type of the magnetic moments ordering of $M_i$ and $M_j$ ions is antiferromagnetic (AF), while if $J_{ij}^s > 0$, the type of the magnetic moments ordering is ferromagnetic (FM).

The room-temperature structural data and ionic radii (CN=6) of Shannon [18] ($r_{V^{4+}}$ = 0.58 Å, $r_{O^{2-}}$ = 1.40 Å, $r_{P^{5+}}$ = 0.38 Å, $r_{S^{6+}}$ = 0.29 Å, $r_{Si^{4+}}$ = 0.40 Å, $r_{Ge^{4+}}$ = 0.53 Å) were used for calculations. Virtually identical results are obtained when similar size of Pauling's ionic radii (CN = 6) [19] are used. To make final conclusions on the magnetic state of a compound it is also necessary to take into account the competition of magnetic interactions.

## 3. Results and discussion

*3.1. The common features of crystal structures and geometrical configurations of $V^{4+}$ ions specific for frustrations of magnetic interactions*

The tetragonal compounds $Zn_2(VO)(PO_4)_2$ [10] (I), $(VO)(H_2PO_4)_2$ [11] (II), $(VO)SiP_2O_8$ [12] (III), $(VO)SO_4$ [13] (IV), $(VO)MoO_4$ [14] (V), $Li_2(VO)SiO_4$ [15] (VI) и $Li_2(VO)GeO_4$ [15] (VII) have very similar structures of sublattices of magnetic ions $V^{4+}$ (figure 1 and table). The structure of $V^{4+}$ ions sublattices is built from linear chains of $V^{4+}$ ions running along the *c*-axis. The distances V-V ($J_1$ couplings) along the chains vary within the limits from 3.983 Å ($(VO)SiP_2O_8$) up to 4.520 Å ($Zn_2(VO)(PO_4)_2$). These chains are located in points of a square lattice parallel to *ab*–plane with the side of the square equal to $a\sqrt{2}/2$.

If one takes the sublattice of $V^{4+}$ ions in the structure $Zn_2(VO)(PO_4)_2$ as an original sublattice, all the others are obtained by shifting by distance $D_1$ (see table 1) along the *c*-axis of chains of $V^{4+}$ ions located, in this case, along one diagonal of a square relatively to chains located along the other diagonal (figure 1). As a result, the plane square lattices of $V^{4+}$ ions existing in the structure of $Zn_2(VO)(PO_4)_2$ turn into the goffered lattices made of distorted squares. The angles ($\angle VVV$) in these distorted squares lie in the range 86° - 90° for all compounds, except $(VO)MoO_4$, in which the angle is equal to 74° (see table 1). The magnetic couplings along the side and diagonal of the square are denoted as $J_2$ and $J_4$, respectively.

It should be emphasized that the $V^{4+}$ ions in the goffered lattices are located at two levels as, for example, in $CaV_4O_9$ [16, 20] and form two plane lattices made of large squares with $J_4$ couplings along the side and $J_8$ couplings along the diagonal of the square (figure 1). The lattices of the large squares are arranged one above the other with a shift along *a*/2 and *b*/2. In this case $J_2$ is a coupling between these lattices. The distance between the lattices of the large squares D1 (table 1) depends on the type of $XO_4$ (X = P, Mo, Si or Ge) groups binding linear chains of $V^{4+}$ ions to one another. For example, in isostructural compounds of $(VO)SO_4$ and $(VO)MoO_4$, an increase of the size of intermediate groups ($SO_4^{2-}$: d(S–O)=1.47 Å; $MoO_4^{2-}$: d(Mo–O)=1.76 Å) results in an increase of D1 from 0.965 Å ($(VO)SO_4$) up to 2.896 Å ($(VO)MoO_4$). Because of the above the goffered lattice in $(VO)MoO_4$ should be more correctly considered as a layer of compressed $V^{4+}$ tetrahedra sharing the edges with four $J_2$ couplings and two $J_4$ couplings. However, in the structure of $(VO)MoO_4$ there is another less distorted square lattice similar to those found in another compounds. This lattice is formed by interactions with the nearest neighbor $J_3$ and the second neighbor $J_4$ (figures 1(e), (f)).

The distance between the goffered lattices D2 considerably exceeds the distance D1 inside the lattice in all compounds except $(VO)MoO_4$. The configuration of the nearest interactions ($J_3$), as well as that of more remote interactions ($J_5$) between $V^{4+}$ ions from neighboring goffered lattices, together with $J_4$ interactions in lattices, represents the layers made of compressed or stretched tetrahedra $V_4$ bound by shared edges.

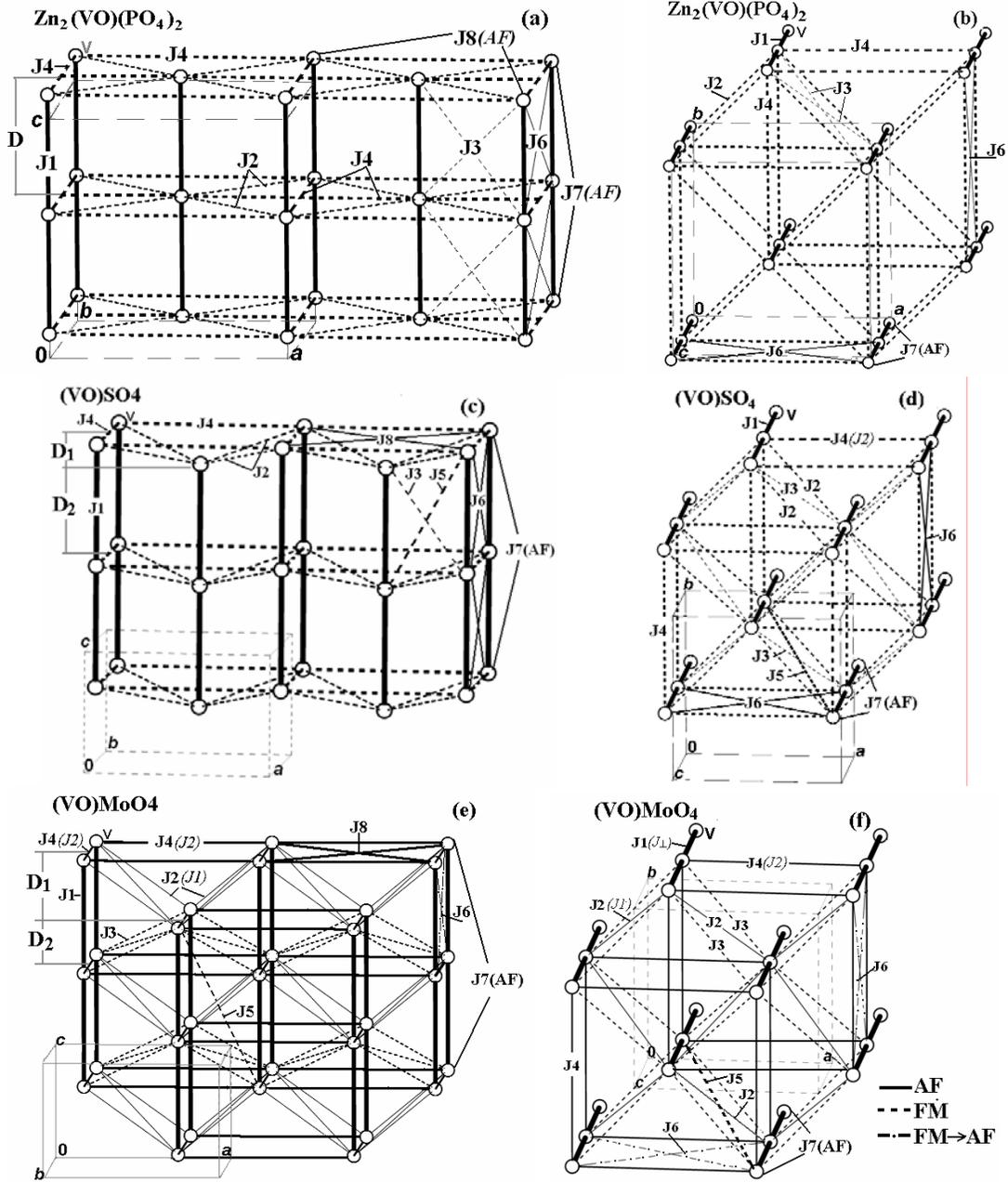

**Figure 1.** A view of $V^{4+}$ sublattices along [010] ((a), (c), (e)) and [001] ((b), (d), (f)) and $J_n$ couplings. The thickness of lines shows the strength of $J_n$ coupling. AF and FM couplings are indicated by solid and dashed lines, respectively. The possible FM→AF transitions are shown by stroke in dashed lines.

Alternatively, the structure of the sublattice of $V^{4+}$ ions can be represented as a network made of crossed diagonal or parallel to diagonal planes of an unit cell with square channels along the *c*-axis (figures 1(b), (d), (f)). The point is that all $V^{4+}$ ions are located in these planes at short distances from each other. They form a rectangular lattice with two non-equivalent nearest-neighbor bonds in the compound I (figure 2(a)) and a distorted triangular lattice with three non-equivalent nearest-neighbor bonds in compounds II–VII (figures 2(b)–(f)).

**Table**. Sign and strength of magnetic interactions ($J_n^s$), calculated on the basis of the structural data and structural parameters.

| | | $Zn_2(VO)(PO_4)_2$ [10] I4cm a = 8.923 Å b = 8.923 Å c = 9.039 Å Z = 4 | $(VO)(H_2PO_4)_2$ [11] P4/ncc a=8.953 Å b=8.953 Å c=7.965 Å Z=4 | $(VO)SiP_2O_8$ [12] P4/ncc a=8.723 Å b=8.723 Å c=8.151 Å Z=4 | $(VO)SO_4$ [13] P4/n a=6.261 Å b=6.261 Å c=4.101 Å Z=2 | $(VO)MoO_4$ [14] P4/n a=6.608 Å b=6.608 Å c=4.265 Å Z=2 | $Li_2(VO)SiO_4$ [15] P4/nmm a = 6.368 Å b = 6.368 Å c = 4.449 Å Z = 2 | $Li_2(VO)GeO_4$ [15] P4/nmm a = 6.487 Å b = 6.487 Å c = 4.517 Å Z=2 |
|---|---|---|---|---|---|---|---|---|
| 1NN | d(V-V) (Å) | 4.520 | 3.983 | 4.076 | 4.101 | 4.265 | 4.449 | 4.517 |
| | $J_1^s \{J\perp\}$ | -0.1653 AF [a] -0.1521 AF | -0.1907 AF [a] -0.1769 AF | -0.1857 AF [a] -0.1728 AF | -0.1811 AF | -0.1687 AF [a] -0.1559 AF | -0.1633 AF [a] -0.1489 AF | -0.1610 AF [a] -0.1471 AF |
| 2NN | d(V-V) (Å) | 6.309 | 6.388 | 6.392 | 4.531 | 5.497 | 4.568 | 4.670 |
| | $J_2^s \{J_1\}$ (Å$^{-1}$) | 0.0147 FM | -0.0161 AF [a] 0.0370 FM | -0.0268 AF [a] -0.0018 AF | 0.0127 FM | -0.0015 AF | 0.0112 FM | 0.0110 FM |
| 3NN | d(V-V) (Å) | 7.761 | 7.063 | 6.617 | 5.426 | 4.869 | 5.817 | 5.857 |
| | $J_3^s \{J_1\}$ (Å$^{-1}$) | 0.0009 FM [a] 0.0015 FM | 0.0149 FM | -0.0051 AF | 0.0035 FM | 0.0088 FM | -0.0008 AF | -0.0017 AF |
| 4NN | d(V-V) (Å) | 8.923 | 8.953 | 8.723 | 6.261 | 6.608 | 6.368 | 6.487 |
| | $J_4^s \{J_2\}$ (Å$^{-1}$) | 0.0325 FM | 0.0042 FM | -0.0081 AF | 0.0272 FM | -0.0150 AF | -0.0174 AF [a] 0.0304 FM | -0.0195 AF |
| 5NN | d(V-V) (Å) | 7.761 | 7.965 | 8.435 | 6.728 | 7.319 | 6.890 | 7.079 |
| | $J_5^s$ (Å$^{-1}$) | 0.0009 FM [a] 0.0015 FM | 0.0100 FM [a] 0.0110 FM | 0.0006 FM [a] 0.0306 FM | 0.0326 FM [a] 0.0345 FM | 0.0204 FM [a] 0.0216 FM | -0.0030 AF [a] -0.0002 AF | -0.0030 AF [a] -0.0006 AF |
| 6NN | d(V-V) (Å) | 10.002 | 9.799 | 9.628 | 7.485 | 7.864 | 7.768 | 7.905 |
| | $J_6^s$ (Å$^{-1}$) | -0.0026 AF | -0.0048 AF | -0.0231 AF [a] -0.0128 AF | -0.0120 AF | -0.0005 AF [a] 0.0001 FM | 0.0028 FM [a] 0.0036 FM | 0.0045 FM [a] 0.0053 FM |
| 7NN | d(V-V) (Å) | 9.039 | 7.965 | 8.151 | 8.202 | 8.530 | 8.896 | 9.034 |
| | $J_7^s$ (Å$^{-1}$) | -0.0261 AF [a] 0.0316 FM | -0.0332 AF [a] 0.0381 FM | -0.0318 AF [a] 0.0358 FM | -0.0314 AF | -0.0290 AF [a] 0.0314 FM | -0.0269 AF [a] 0.0239 FM | -0.0261 AF [a] 0.0218 FM |
| 8NN | d(V-V) (Å) | 12.619 | 12.661 | 12.336 | 8.854 | 9.345 | 9.006 | 9.174 |
| | $J_8^s$ (Å$^{-1}$) | -0.0305 AF | -0.0192 AF | -0.0106 AF | -0.0032 AF | -0.0252 AF | -0.0082 AF [a] 0.0492 FM | -0.0081 AF |
| D1 (Å) | | 0 | 0.852 | 1.680 | 0.965 | 2.896 | 0.766 | 0.875 |
| D2 (Å) | | 4.520 | 3.131 | 2.396 | 3.136 | 1.369 | 3.683 | 3.642 |
| Angle ∠VVV between 2NN (deg) | | 90 | 88.98 | 86.05 | 87.40 | 73.89 | 88.39 | 87.99 |
| Angle ∠VVV between 3NN (deg) | | 70.18 | 78.67 | 82.46 | 70.48 | 85.47 | 66.37 | 67.26 |
| Angle ∠VVV between 5NN (deg) | | 70.18 | 68.39 | 79.68 | 55.48 | 53.67 | 55.05 | 54.54 |
| Linear chains along the c-axis | $J_1^s/J_7^s \{J_\perp/J_7\}$ | 6.33 (AF/AF) | 5.74 (AF/AF) | 5.84 (AF/AF) | 5.77 (AF/AF) | 5.82 (AF/AF) | 6.07 (AF/AF) | 6.17 (AF/AF) |
| Lattice of smaller distorted squares | $J_4^s/J_2^s \{J_2/J_1\}$ | 2.21 (FM/FM) | -0.26 (FM/AF) | 0.30 (AF/AF) | 2.14 (FM/FM) | 10.0 (AF/AF) | -1.55 (AF/FM) | -1.77 (AF/FM) |
| Lattice of larger squares | $J_8^s/J_4^s$ | -0.93 (AF/FM) | -4.57 (FM/AF) | 1.31 AF/AF | -0.12 (AF/FM) | 1.68 (AF/AF) | 0.47 (AF/AF) | 0.42 (AF/AF) |
| Edge sharing distorted tetrahedral lattices | $J_4^s/J_3^s$ | 36.1 (FM/FM) | 0.28 (FM/FM) | 1.59 (AF/AF) | 7.77 (FM/FM) | -1.71 (AF/FM) | 21.8 (AF/AF) | 11.47 (AF/AF) |
| Edge sharing distorted tetrahedral lattices | $J_4^s/J_5^s$ | 36.1 (FM/FM) | 0.04 (FM/FM) | -0.27 (AF/FM) | 0.83 (FM/FM) | -0.74 (AF/FM) | 5.80 (AF/AF) | 6.50 (AF/AF) |

[a] During calculation of $J^s$ coupling the contribution from an intermediate ion located in critical position (a) was not taken into account.

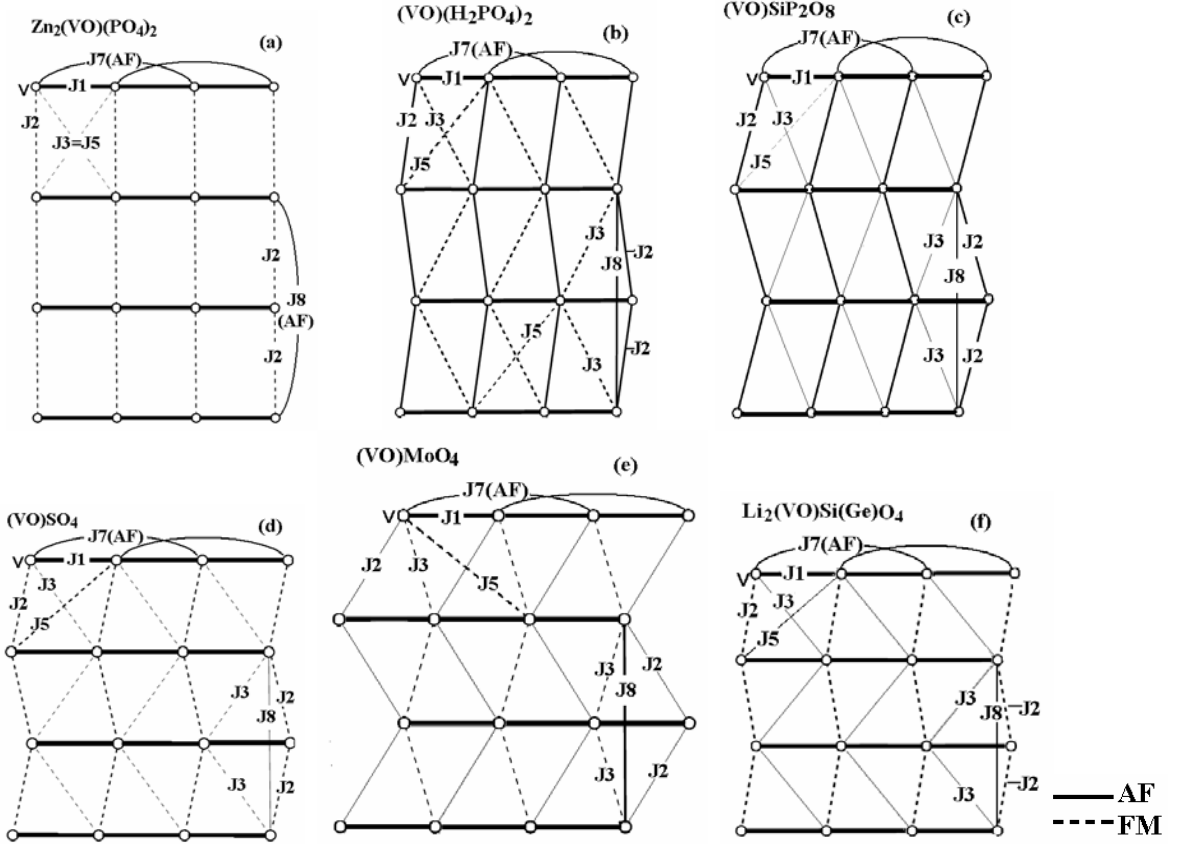

**Figure 2.** Rectangular (a) and triangular ((b) – (f)) lattices and $J_n$ couplings. The thickness of lines shows the strength of $J_n$ coupling.

A basic element of magnetic structure in a rectangular lattice is the rectangle with $J_1$ and $J_2$ couplings, and in a triangular lattice, the triangle with $J_1$, $J_2$ and $J_3$ couplings. The triangular lattice can be also considered as a four-cornered, in which the basic element is parallelogram.

The couplings $J_5$, $J_7$ and $J_8$ are the next-to-nearest neighbor couplings in diagonal planes (figure 2) while $J_4$ and $J_6$ couplings pass through the channels between these planes (figure 1).

The distances d(V–V) between magnetic ions for $J_1$–$J_8$ couplings are given in table 1.

Thus, in crystal structures discussed above the following specific geometrical configurations of $V^{4+}$ ions (figure 1 and 2), which can result in geometric magnetic frustrations, were found:

- linear chains along the *c*-axis with nearest-neighbor $J_1$ and next-nearest-neighbor $J_7$ intrachain couplings;
- rectangular lattices (in I) with non-equivalent nearest-neighbor $J_1$ and $J_2$ couplings and triangular lattice (in II-VII) with non-equivalent nearest-neighbor $J_1$, $J_2$ and $J_3$ couplings and next-nearest-neighbor couplings $J_5$, $J_7$ and $J_8$;
- square lattice or distorted square lattice (smaller squares) with nearest-neighbor $J_2$ and next-nearest-neighbor (diagonal) $J_4$ couplings ($J_3$ and $J_4$ in (VO)MoO$_4$);
- square lattice (larger squares) with nearest-neighbor $J_4$ and next-nearest-neighbor (diagonal) $J_8$ couplings;

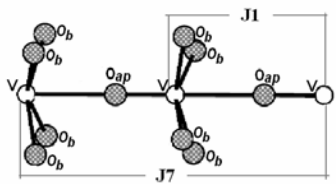

**Figure 3.** Arrangement of intermediate oxygen ions in space of $J_1$ and $J_7$ interactions.

- linear or slightly zigzag (bent) chains along diagonals in the plane *ab* with nearest-neighbor $J_2$ ($J_3$ in (VO)MoO$_4$)) and next-nearest-neighbor $J_8$ intrachain couplings;
- edge sharing distorted tetrahedral lattices with four $J_3$ and two $J_4$ couplings;
- edge sharing distorted tetrahedral lattices with four $J_5$ and two $J_4$ couplings.

*3.2. Characterization of magnetic interactions and the origin of the differences between the compounds*

The strength of magnetic interactions and the ordering type of the ions magnetic moments in Zn$_2$(VO)(PO$_4$)$_2$ (I), (VO)(H$_2$PO$_4$)$_2$ (II), (VO)SiP$_2$O$_8$ (III), (VO)SO$_4$ (IV), (VO)MoO$_4$ (V), Li$_2$(VO)SiO$_4$ (VI) and Li$_2$(VO)GeO$_4$ (VII) determined by a crystal chemistry method with using the program "MagInter" are given in table 1.

The composition of intermediate ions and their geometrical arrangement in the space of interaction between magnetic V$^{4+}$ ions along the *c*-axis ($J_1$ and $J_7$ couplings) are virtually identical in all compounds I–VII. However, in other interactions ($J_2$–$J_6$ and $J_8$) such similarity is observed only within the limits of one group of compounds I–III or IV–VII. Besides, there are some displacements of intermediate ions or their substitution by other ions even in isostructural compounds.

*3.2.1. Magnetic interactions in linear chains along the c-axis.* The strong antiferromagnetic nearest-neighbor $J_1$ interactions in linear chains along the *c*-axis (figure 3, table 1) are dominant in all compounds under consideration. The contribution (from -0.161 Å$^{-1}$ up to -0.191 Å$^{-1}$) to the AF component of this interaction results from the apical oxygen ion (O$_{ap}$), which is located on a direct line connecting V ions and divides this line into nonequivalent parts *l* and *l'* (*l'*/*l*<2.0). The distance $h$(O$_{ap}$) from the centre of O$_{ap}$ ion up to a straight line connecting V$^{4+}$ ions is equal to zero.

Besides the O$_{ap}$ ion, in all compounds (except (VO)SO$_4$) four oxygen ions O$_b$ from four tetrahedral groups of XO$_4$ (X = P, Mo, Si or Ge), which form the basal plane of VO$_5$ pyramid, are included into the $J_1$ space interaction (a cylinder with radius 0.58 Å and length from 3.983 Å up to 4.520 Å). These ions are responsible for an occurrence of comparatively small contribution to the FM component of interaction that, on summing up the contributions with different signs, reduces the size of $J_1$ just a little (see table 1). In addition, they are located near the border of interaction space (critical point "a"; see section 3 in [16]), since the distances $h$(O$_b$) from the centre of O$_b$ ions to the V-V line are close to the critical distance ($h_c$(O)) that is equal to the sum of radii of V$^{4+}$ and O$^{2-}$ ions (1.98 Å). An increase of $h(O_b)$ just by 0.05 Å (in all compounds except Li$_2$(VO)SiO$_4$ and Li$_2$(VO)GeO$_4$, where $h(O_b)$ should be increased up to ~0.1 Å) displaces them beyond the interaction space and, therefore, excludes their FM contribution. It could occur at reduction of the distance V-V with temperature decrease, since the distance $h$(O$_b$) increases from 1.855 Å up to 2.00 Å with decrease of d(V-V) (parameter *c*) from 4.517 Å down to 4.101 Å for a number of compounds: Li$_2$(VO)GeO$_4$, Li$_2$(VO)SiO$_4$, (VO)MoO$_4$ and (VO)SO$_4$. Besides, it was experimentally proved [8, 9] that in (VO)MoO$_4$ compounds the parameter *c* reduces at cooling that is accompanied by a removal of basal O$_b$ ions from the VO-VO chain.

The dominant intrachain AF $J_1$ interaction could determine the structure of the magnetic system of these compounds as a strong one-dimensional antiferromagnet in the absence of competition from the nearest-neighbor $J_1$ interactions with the next-nearest-neighbor $J_7$ (d(V-V)=2$c$) interactions in a chain. The $J_7$ interactions are approximately six times weaker ($J_1/J_7$ = 5.74–6.33) than $J_1$ interactions and are antiferromagnetic, if one does not take into account FM contributions from eight basal oxygen ions, $O_b$, located in critical positions (figure 3, table 1).

It should be emphasized that, according to our calculations, the contribution from eight ions $O_b$ into the FM component of $J_7$ interaction exceeds by a factor 2 AF contribution from two $O_{ap}$ ions and, hence, taking into account these contributions leads to the FM-type of $J_7$ interaction. However, the comparison of results obtained with using a crystal chemical method with the data of other methods performed in [16] has shown that if the intermediate ions are arranged in critical positions close (~0.1Å) to the surface of a cylinder bounding the space between magnetic ions (critical point "a"), they, as a rule, do not participate in emergence of magnetic interactions. This is related to several factors: first, the difference of conditions (temperature, pressure) of the structural and magnetic properties of compounds studied; second, the accuracy of structural data. Nevertheless, it should be mentioned that even insignificant distortions of structure can result in total or partial involvement of $O_b$ ions in the interaction space and the emergence of the $J_7$ interaction FM component and, finally, in magnetic anomalies in a VO-VO chain.

*3.2.2. Square lattice with nearest-neighbour coupling $J_2$ and next-nearest-neighbor (diagonal) coupling $J_4$.* The layers containing $V^{4+}$ square lattice with the nearest-neighbor coupling $J_2$ and next-nearest-neighbor (diagonal) coupling $J_4$ are formed by the $VO_5$ square pyramids sharing corners with $XO_4$ (X = P, Mo, Si or Ge) tetrahedron. However, the number and location of $XO_4$ tetrahedra differ in compounds of the first $Zn_2(VO)(PO_4)_2$ (I), $(VO)(H_2PO_4)_2$ (II), $(VO)SiP_2O_8$ (III), and the second $(VO)SO_4$ (IV), $(VO)MoO_4$ (V), $Li_2(VO)SiO_4$ (VI), $Li_2(VO)GeO_4$ (VII)) groups. The structures of layers for typical representatives of these groups, $(VO)SiP_2O_8$ and $(VO)SO_4$, are shown in figure. 4(a) and (b). In compounds I–III the $XO_4$ tetrahedra are located in the middles of the sides of squares of the $V^{4+}$ lattice while each of them shares two corners only with two $VO_5$ pyramids. In the compounds IV–VII the number of $XO_4$ per $V^{4+}$ ion is two times lower, which forces the remaining tetrahedra to move to the square center and use all the oxygen atoms to bind with $V^{4+}$ ions. As a result, a compression of the $V^{4+}$ square lattice takes place.

The group $PO_4$ located between the vanadium ions controls the spin orientation and strength of $J_2$ interaction in compounds I–III (figure 4(a)). The two oxygen ions of this group located near the middle of the V–V line initiate the contribution to the AF component of $J_2$ interaction; the other two oxygen ions and the phosphorus ion located near the surface of the cylinder bounding the space between magnetic ions can initiate a substantial contribution to the FM component of this interaction in case of entering the interaction space. The oxygen atoms from other $PO_4$ groups are removed from the middle to $V^{4+}$ ions and, therefore, can initiate only an insignificant contribution to the FM component of the interaction.

The strongest $J_2$ interaction is in $(VO)SiP_2O_8$ ($J_2^s$ = -0.0268 Å$^{-1}$) (figure 4(c)), since $PO_4$ group is located farther ($h$(P)=0.876 Å) from the line V–V than in $Zn_2(VO)(PO_4)_2$ ($h$(P)=0.736 Å) and $(VO)(H_2PO_4)_2$ ($h$(P)=0.524 Å). As a result, there are only two oxygen ions ($h$(O)=0.164 Å) in the space of interaction that initiate the contribution to the AF component of $J_2$ interaction. The phosphorus ions are in critical position "a" and their FM contribution ($j^s$(P) = 0.0243 Å$^{-1}$), apparently, should not be taken into account.

In the compound $(VO)(H_2PO_4)_2$ the strength of the AF component of $J_2$ interaction ($J_2^s$ = -0.0161 Å$^{-1}$) is substantially lower than in $(VO)SiP_2O_8$ because of the approach of $PO_4$ group to the V–V line. As a result, the phosphorus ion enters the interaction space and initiates the FM contribution equal to 0.0071 Å$^{-1}$, whereas two ions of oxygen are slightly ($h$(O)=0.325 Å) removed and reduce the contribution to the AF component of the $J_2$ interaction up to -0.0232 Å$^{-1}$. One should be aware that two other oxygen ions of the $PO_4$ group also cross (at 0.06 Å) the border of the interaction space (critical position 'a') while each of them can initiate a substantial contribution to the FM component of the interaction ($j^s$ (O) = 0.0258 Å$^{-1}$) and cause the AF→FM transition in $J_2$ interaction.

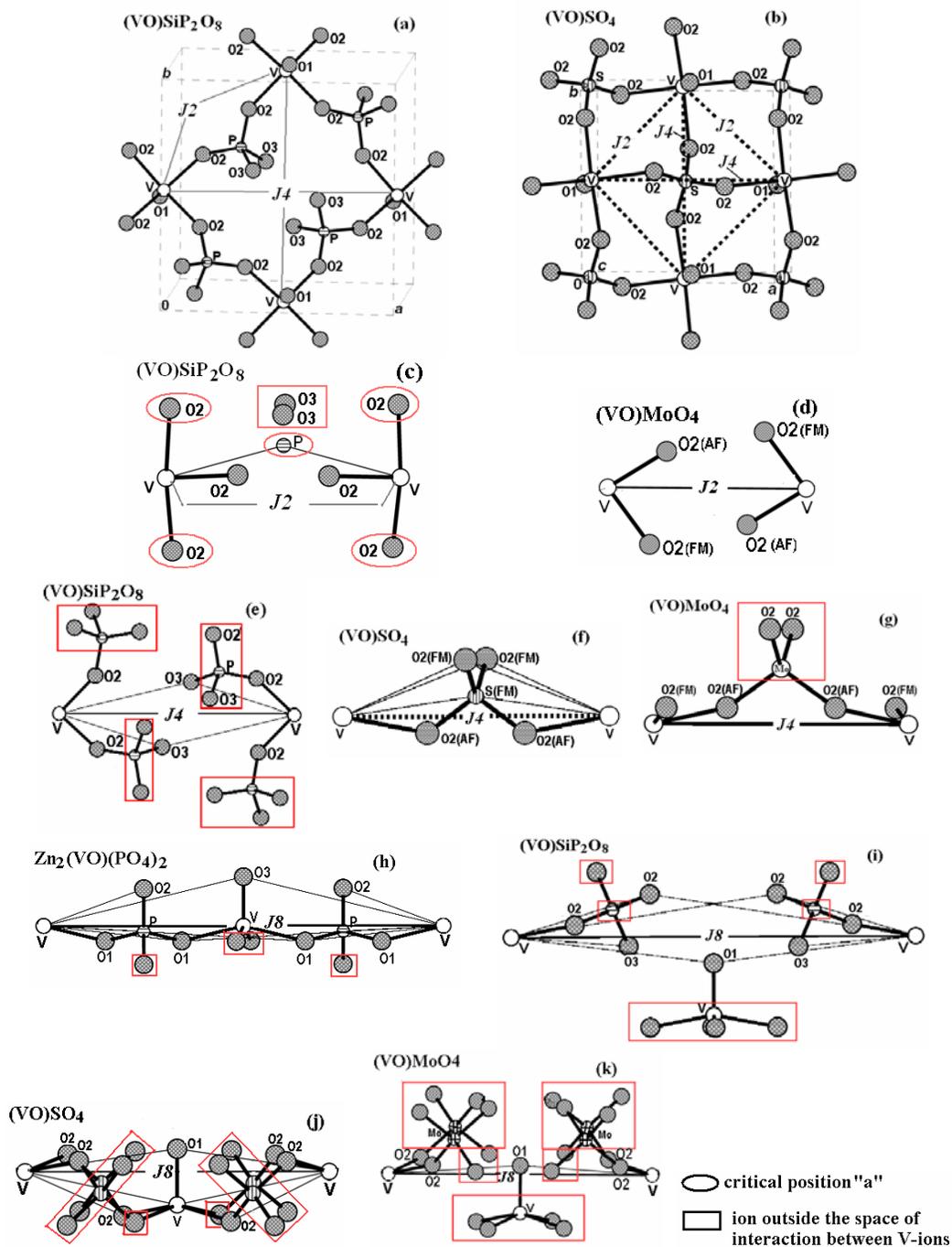

**Figure 4.** Structure of layers of $VO_5$ square pyramids sharing corners with $XO_4$ (X = P, Mo, Si or Ge), in I-III (a) and IV-VII (b) compounds. The arrangement of intermediate ions in space of $J_2$ ((c)-(d)), $J_4$ ((e)-(g)), and $J_8$ ((h)-(k)) interactions.

This very case is characteristic for $Zn_2(VO)(PO_4)_2$ ($J_2^s = 0.0147$ Å$^{-1}$, FM) where one of the oxygen ions moves deep by 0.26 Å into the interaction space. As a result, the contributions to the FM component of interaction from this oxygen ion ($j^s$(O) = 0.0161 Å$^{-1}$) and phosphorus ion ($j^s$(P) = 0.0178 Å$^{-1}$) exceeded (by 0.0147 Å$^{-1}$) the value of contributions from two oxygen ions to the AF component of interaction ($j^s$(O) = -0.0096 X2 Å$^{-1}$).

In the second group of compounds the $J_2$ coupling emerges mainly from four basal oxygen ions, $O_b$, of two $VO_5$ pyramids or two $XO_4$ groups (figure 4(d)). The distances $h(O_b)$ from the centre of these ions $O_b$ to the V–V line get closer (1.22–1.70 Å) to the radius of oxygen ion ($r_{O^{2-}}$ = 1.40 Å), i.e. the network of $O_b$ ions is located near the V-V line. In [16] we have shown that such a location of intermediate ions is critical (critical point "b"). In this case a weak AF interaction emerges on insignificant reduction of $h(O_b)$ (the overlapping of the bond line ($h(O_b) < r_{O^{2-}}$) by $O_b$ ion) while a weak FM interaction emerges on increase of $h(O_b)$ (formation of a gap between $O_b$ ion and the line of bond V–V ($h(O_b) > r_{O^{2-}}$)). That is why the character of $J_2$ interaction in compounds IV–VII varies from weak antiferromagnetic in $(VO)MoO_4$ ($J_2$ = -0.0015 Å$^{-1}$) to ferromagnetic ($J_2$ = 0.0110 – 0.0127 Å$^{-1}$) for all other compounds of the second group.

The $J_4$ interactions in the compounds I–III are formed mainly due to oxygen ions from two groups of $PO_4$, which are located in the central one-third of space ($l'/l < 2.0$) between $V^{4+}$ ions (figure 4(e)). In $Zn_2(VO)(PO_4)_2$ ($J_4$ = 0.325 Å$^{-1}$ (FM)) two oxygen ions from each group enter the interaction space and initiate the emergence of the contributions to the FM component of $J_4$ interaction. In the compounds $(VO)(H_2PO_4)_2$ ($J_4$ = 0.042 Å$^{-1}$ (FM)) and $(VO)SiP_2O_8$ ($J_4$ = -0.081 Å$^{-1}$ (AF)) only one oxygen ion per compound remains in the $J_4$ interaction space due to $PO_4$ turning. These oxygen ions are in a critical position ($h$(O) = 1.25–1.48 Å; critical point "b") and initiate the contribution to the FM component (in $(VO)(H_2PO_4)$) and to the AF component (in $(VO)SiP_2O_8$) of $J_4$ interaction. The contributions to the AF and FM components of $J_4$ interaction from the four basal oxygen ions of $VO_5$ pyramids are insignificant and suppress each other.

As in the case of $J_2$ couplings in compounds I-III, the $XO_4$ group located between $V^{4+}$ ions determines the sign and strength of $J_4$ interactions in compounds IV–VII. In $(VO)SO_4$ the FM $J_4$ coupling ($J_4$ = 0.0272 Å$^{-1}$) emerges under the influence of all ions of $SO_4$ group (figure 4(f)). Two oxygen ions of this group make the contribution ($j^s$(O)= -0.0111 X 2 Å$^{-1}$) to the AF component while two other oxygen ions and the sulfur ion make the contribution ($j^s$(O)= 0.0198 X 2 Å$^{-1}$, $j^s$(S)= 0.0098 Å$^{-1}$) to the FM component of $J_4$ interaction. In isostructural compound $(VO)MoO_4$ ($J_4$ = -0.0150 Å$^{-1}$) within the space limits of $J_4$ coupling, there are only two oxygen ions of the $MoO_4$ group, and they cause the AF spins ordering (figure 4(g)). In $Li_2(VO)SiO_4$ and $Li_2(VO)GeO_4$ the Si(Ge) ions are the nearest to the V–V line and, together with two oxygen ions, they initiate the AF $J_4$ interaction ($J_4$ = -0.0174(-0.0195) Å$^{-1}$). However, in Si-compounds this interaction is unstable because of the two other oxygen ions located in critical position "a" ($h$(O)=1.886 Å). The oxygen ions are capable of making a substantial contribution ($j^s$(O) = 0.0258 X 2 Å$^{-1}$) to the FM component of $J_4$ interaction and cause a transition AF→FM. In the Ge compound, in contrast to Si-compound, these ions have no effect on formation of the $J_4$ interaction, since they leave the space of $J_4$ interaction ($h$(O)=2.032 Å)) because of the greater size of the $GeO_4$ group (d(Ge–O)=1.74 Å) as compared to the $SiO_4$ group ((d(Si–O)=1.633 Å) and smaller (on 0.055 Å; $h$(Ge) = 0.438 Å and $h$(Si)=0.383 Å) shifting to the V–V line.

The $J_8$ interaction ($J_8$ = -0.0032 – -0.0305 Å$^{-1}$; d(V–V) ~ 2$NN$) along the chain with the nearest-neighbor $J_2$ is of AF-type in all compounds under study (figures 4(h)–(k)). The AF character of this interaction in compounds I–II is determined mainly by two basal oxygen ions while in compounds III–VII it is determined by one apical oxygen ion from the intermediate $VO_5$ pyramid, which is located near the middle ($l'/l \leq 2.0$; $h$(O) = 0.31–0.86 Å) of the line of V–V interaction. The $V^{4+}$ ion enters the space of $J_8$ interaction in all compounds, except for two where $h(V) > 2 r_{V^{4+}}$ [(VO)SiP$_2$O$_8$ ($h(V)$=1.679 Å) and (VO)MoO$_4$ ($h(V)$=1.369 Å)], and makes the FM contribution in compounds II, IV, VI and VII and the AF contribution in Zn$_2$(VO)(PO$_4$)$_2$. The contribution from V$^{4+}$ ion is several-fold less in absolute value than the contribution from oxygen ions.

It is necessary to emphasize that only in one (Li$_2$(VO)SiO$_4$ and Li$_2$(VO)GeO$_4$) from two pairs of isostructural compounds (IV–V and VI–VII) are the characteristics of all appropriate magnetic interactions are close to each other. The difference in magnetic interactions in Li$_2$(VO)SiO$_4$ from those in Li$_2$(VO)GeO$_4$ consists in the possibility of the transition AF→FM in $J_4$ and $J_8$ couplings of Si-compounds. The interaction $J_8$ in Li$_2$(VO)SiO$_4$ can undergo a transition like the interaction $J_4$, since four O$_b$ ions of the intermediate VO$_5$ pyramid are present additionally in the interaction space in critical position "a", and their contribution to the FM component of interaction is four times higher than the AF contribution from the O$_{ap}$ ion.

*3.2.3. Magnetic interactions between layers.* The nearest interactions $J_3$ (figures 5(a)-(c) and table 1) between layers in all compounds are weak (AF in compounds III, VI and VII, and FM in compounds I, II, IV and V) and unstable for two reasons. First, most of intermediate oxygen ions are in critical position "b" ($h(O) \approx r_{O^{2-}}$; see section 3 in [16]) and, second, the sum of contributions $j_O^s$ to the AF component of interaction becomes closer to the sum of contributions to the FM component of interaction (critical point "d"; see section 3 in [16]). As a result, slight displacement of even one of the intermediate ions can result in a loss of interaction or reorientation of spins.

The more remote $J_5$ interactions (table 1) differ considerably from each other both in the value and in the ordering type of magnetic moments. If one does not take into account the contributions from ions located in critical positions, the interactions $J_5$ are weak AF in Li$_2$(VO)Si(Ge)O$_4$ ($J_5$= -0.003(-0.003 Å$^{-1}$), weak FM in (VO)SiP2O8 ($J_5$= 0.0006 Å$^{-1}$) and rather strong FM interactions in other compounds ($J_5$ = 0.010–0.033 Å$^{-1}$). In (VO)(H$_2$PO$_4$)$_2$ and (VO)SiP$_2$O$_8$ (figure 5(d)) small FM contributions from two ions O$_{ap}$ of chains VO–VO are approximately equal to AF contributions from the two O$_b$ ions and suppress each other. However, in the central part of $J_5$ couplings space in (VO)(H$_2$PO$_4$)$_2$, there are two oxygen ions from two groups PO$_4$ ($h(O) = 1.57$ Å; $l'/l \leq 2.0$), whose contribution to the FM component of interaction ($j_O^s$ = 0.0054 X 2 Å$^{-1}$) is crucial. In (VO)SiP$_2$O$_8$ these oxygen ions move from line V-V at the distance $h(O) = 1.88$ Å, whereas the two other oxygen ions from these PO$_4$ groups cross the border ($h(O) = 1.93$ Å; $l'/l \leq 2.0$) and enter the space of interaction. Each of these ions can initiate a large FM contribution equal 0.0150 Å$^{-1}$. However, all of these ions are there in critical positions "a", since their $h(O)$ get close to the critical value ($h_c(O) = 1.98$ Å). In (VO)S(Mo)O$_4$ ($J_5$ = 0.0204 Å$^{-1}$ (FM)) (figure 5(e)) and Li$_2$(VO)Si(Ge)O$_4$ (figure 5(f)) two O$_{ap}$ and two or four O$_b$ ions initiate small AF (-0.003 Å$^{-1}$) and FM (0.002–0.003 Å$^{-1}$) contributions to the $J_5$ interaction. However, O$_b$ ions are in critical positions "a" ($h(O_b) = 1.85$–1.93 Å). Besides, in the $J_5$ couplings space in (VO)S(Mo)O$_4$, there are additionally two oxygen ions, which make a large FM contribution (0.033(0.021) Å$^{-1}$) that results in rather strong FM $J_5$ interaction, unlike Li$_2$(VO)Si(Ge)O$_4$.

Relatively strong AF $J_6$ interactions in planes *ac* and *bc* along diagonals of a rectangle with sides 1NN and 4NN are present only in compounds (VO)SiP$_2$O$_8$ (-0.0231 Å$^{-1}$) and (VO)SO$_4$ (-0.0120 Å$^{-1}$) (figures 5(g), (h), (i)). They emerge mainly due to the effect of two oxygen ions of P(S)O$_4$ group, which get closer to the line V–V. In other compounds the ions included in the $J_6$ coupling space make small contributions. In this case in compounds I, II, V the sum of AF contributions exceeds the sums of FM contributions, while in compounds VI and VII the opposite is the case.

Thus, we have shown that XO$_4$ groups (figures 4, 5) determine the sign and strength of the majority of magnetic interactions. The ions of the same XO$_4$ group are intermediate ions involved simultaneously in several interactions and are quite often located in the space of interaction between magnetic ions in critical positions "a", "b" or "d" (see section 3 in [16]). The X–O bonds in XO$_4$ groups are "rigid", therefore, the effect of temperature or pressure can lead to displacement or rotation of a group as a whole unit. This can result not only in drastic changes of strength of several magnetic interactions at once, but also in the transition AF–FM even without the compounds symmetry reduction.

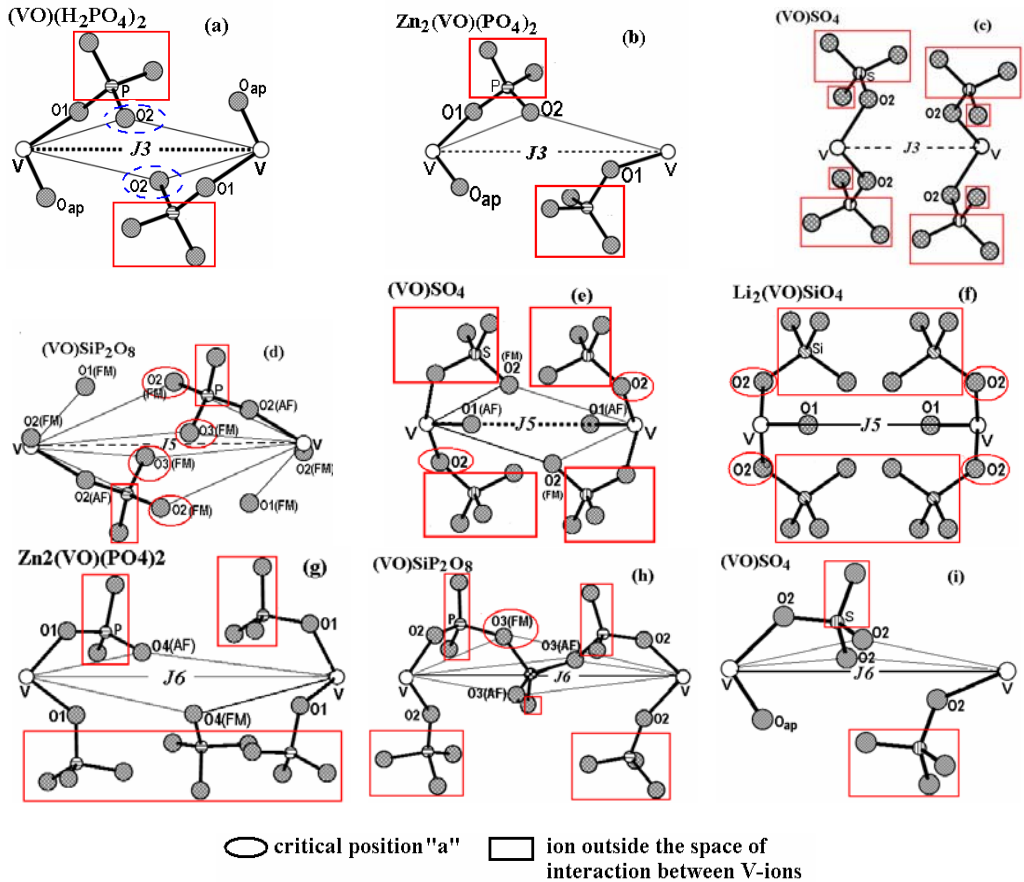

**Figure 5.** The arrangement of intermediate ions in the space of $J_3$ (a)-(c), $J_5$ (d)-(f), and $J_6$ (g)-(i) interactions.

The calculation of the sign and strength of magnetic interactions described above was based on the structural data obtained at room-temperature. Use of these data for estimation of characteristics of magnetic interactions at low temperature can result in errors mainly in cases when intermediate ions are located in critical positions. That is why we present in the table 1 two values of magnetic interactions: with and without taking into account the contributions from intermediate ions located in critical positions close (~0.1Å) to the surface of the cylinder bounding region of the space between the magnetic ions (critical point "a"). Additional errors may be associated with estimation of the size of magnetic ion $V^{4+}$ and intermediate ions. Nevertheless, as was shown in [16], the results have enabled us to establish the main features of magnetic interactions in these seven compounds.

*3.3. Competition of magnetic interactions on specific geometrical configurations of $V4^+$ ions*

Our estimation of magnetic interactions shows that in all compounds I–VII the antiferromagnetic nearest-neighbor $J_1$ interactions in linear chains along the *c*-axis are at least six times stronger than other $J_2$–$J_8$ interactions. However, these interactions compete with weaker $J_7$ ( $J_7^s/J_1^s \cong 1/6$) AF next-nearest-neighbor interactions in a chain. For nearest- and next-nearest-neighbor interactions in antiferromagnetic chain, instability against spontaneous dimerization for $J_{NNN} > J_{NNN}^c \cong \frac{1}{6} J_{NN}$ was found [21].

One can assume that, in addition to the competition between $J_1$ and $J_7$ interactions in chains along the $c$-axis, a competition between other interactions in these systems takes place. Let us consider the competition between interactions on the basis of representation of the structure of these compounds as crossed diagonal planes of a unit cell. As shown above, these planes comprise rectangular lattices with two non-equivalent nearest-neighbor bonds ($J_1 \neq J_2$) in $Zn_2(VO)(PO_4)_2$ and distorted triangular lattices with three non-equivalent nearest-neighbor bonds ($J_1 \neq J_2 \neq J_3$) in other compounds (figure 2). Determination of competition between interactions $J_1$, $J_2$ and $J_3$ in rectangulars and triangles, which are basic elements of these lattices, is of great difficulty because of non-equivalence of strengths of interactions and superposition of competitions by additional next-nearest-neighbor interactions. Based only on the signs of interactions, which we calculated, the rectangles with FM $J_3$ interactions (along diagonals) and AF $J_1$ and FM $J_2$ interactions (along the sides) ($J_1 = -11J_2 = -184J_3$) in a rectangular lattice of $Zn_2(VO)(PO_4)_2$ (figure 2(a)) can be geometrically frustrated. The distorted triangular lattices can be geometrically frustrated only in two compounds: in $(VO)SiP_2O_8$ (figure 2(b)), where all three $J_1$, $J_2$ and $J_3$ interactions in triangle are antiferromagnetic ($J1 = 6.9J_2 = 36.4J_3$), and in $(VO)SO_4$ (figure 2(d)) with one AF $J_1$ interaction and two FM $J_2$ and $J_3$ interactions ($J_1 = -14.3J_2 = -51.7J_3$). In $(VO)MoO_4$ frustration is possible in another triangle with one AF $J_1$ interaction and two FM $J_3$ and $J_5$ interactions ($J_1 = -19.2J_3 = -8.3J_5$) (figure 2(e)).

Beside frustrations in basic elements of lattices, geometrical frustrations exist in linear chains between the FM nearest-neighbor $J_2$ and the AF next-nearest-neighbor $J_8$ ($J_2^s/J_8^s = -0.48$) in $Zn_2(VO)(PO_4)_2$ (figure 2(a)). In other compounds the competition between AF interaction $J_8$ and interactions $J_2$ ($\left|J_2^s/J_8^s\right| = 0.1$–4.0) and $J_3$ ($\left|J_3^s/J_8^s\right| = 0.1$–1.1) in two triangles ($J_2J_2J_8$ and $J_3J_3J_8$) is inevitable in tetragonal symmetry at any signs of $J_2$ and $J_3$ interactions (figure 2(b)-(f)). These frustrations can disappear through lattice deformation from tetragonal symmetry into orthorhombic symmetry.

These lattices can be represented differently as "rectangular" (in $Zn_2(VO)(PO_4)_2$) and "triangular" (in II–VII) n-leg ladders (figure 2) with strong AF legs ($J_1$) and weak AF or FM rungs ($J_2$) ($\left|J_2^s/J_1^s\right| = 0.01$–0.09). The additional next-nearest-neighbor couplings also are weak: in legs ($J_7$) they are antiferromagnetic ($J_7^s/J_1^s = 0.16$–0.17), and both diagonal couplings ($J_3$, $J_5$) can be of any kind ($\left|J_3^s/J_1^s\right| = 0.005$–0.08; $\left|J_5^s/J_1^s\right| = 0.005$–0.18; $\left|J_2^s/J_3^s\right| = 0.17$–16.3; $\left|J_2^s/J_5^s\right| = 0.07$–44.6). Determination of the effect of additional next-to-nearest-neighbor interactions in these frustrated spin ladders is also a difficult problem.

Recently [1–7], the magnetic structure of $(VO)MoO_4$, $Li_2(VO)SiO_4$ and $Li_2(VO)GeO_4$ compounds has been considered as an antiferromagnetic square lattice with nearest-neighbor $J_2$ couplings along the sides of the square and next-nearest neighbor $J_4$ couplings along the diagonal of the square (figure 1(b), (d), (f)). The nearest interactions between square lattices correspond to the nearest-neighbor $J_1$ interactions in linear chains along the $c$-axis. In the literature a different notation is used that is related to ours through $J_1 \rightarrow J_\perp$, $J_2 \rightarrow J_1$ and $J_4 \rightarrow J_2$ (fig. 1f). (It should be mentioned that $J_1\{J_\perp\}$ and $J_2\{J_1\}$ of interactions are located in diagonal planes of unit cell, and $J_4$ interactions connect these crossed planes through a square channel.) Originally [8, 9] the conclusion was made that $(VO)MoO_4$ was a one-dimensional antiferromagnet. Later [1, 2] all three compounds: $(VO)MoO_4$, $Li_2(VO)SiO_4$ and $Li_2(VO)GeO_4$ were considered as two-dimensional antiferromagnets with very weak interplanar couplings. This has allowed using them as prototypes of frustrated two-dimensional antiferromahnet on square-lattice for studies of the role of frustrating interactions in low-dimensional systems [3–7].

Estimation of the nearest-neighbor coupling $J_2\{J_1\}$ and second-neighbor (diagonal) coupling $J_4\{J_2\}$ is of special interest, since the frustration ratio $\alpha = J_4/J_2\{J_2/J_1\}$ is used for making phase diagrams and determination of the systems ground states and phase transitions. However, the results of determination of the α ratio vary substantially between authors. The value α found by Carretta et al. [1] as $\alpha > 1/2$ for $Li_2(VO)(Si,Ge)O_4$ is much less than $\alpha \cong 1 - 4$, obtained by the same authors later [3]. Rosner et al. [4, 5] increased the value α up to 5 in $Li_2(VO)GeO_4$ and up to 12 in $Li_2(VO)SiO_4$. There is also no clarity in regard to the α ratio value in $(VO)MoO_4$. Carretta et al. [2] concluded that nearest- and next-nearest-neighbor interactions in a square lattice were approximately equal ($\alpha \cong 1$), whereas Bombardi et al. [6] argued that the value of the nearest-neighbor interaction was significantly higher than that of the next-to-nearest-neighbor interaction ($\alpha << 1/2$).

The values of α ratio (α = |1.6–1.8|) between next-nearest-neighbor ($J_4$) and nearest-neighbor ($J_2$ in $Li_2(VO)(Si,Ge)O_4$ and $J_3$ in $(VO)MoO_4$ (see section 3.2)) interactions we obtained are well within the permissible limits ($1 \leq \alpha \leq 4$) [2, 3]. However, as was shown above, the columns along c-axis are formed not only by $J_2$ interactions, but also by $J_3$ interactions, which can be presented as distorted square lattices (figure 1(b), 1(d), 1(f)). The absolute value of the ratio of $J_4$ to the sum of values $J_2$ and $J_3$ also lies within these limits (2.1 in $Li_2(VO)SiO_4$, 1.67 in $Li_2(VO)GeO_4$ and 2.1 in $(VO)MoO_4$). It is interesting that we obtained large values of α for the ratios $J_4/J_3(J_4/J_5)$ [22(6) in $Li_2(VO)SiO_4$ and 11(7) in $Li_2(VO)GeO_4$] and for respective $J_4/J_2$ ratio in (VO)MoO4, which is equal to 10.

According to our data, the value of α ratio in $(VO)SO_4$ ($J_4/J_2$ = 2.13) and $Zn_2(VO)(PO_4)_2$ ($J_4/J_2$ = 2.21) is slightly greater than in $(VO)MoO_4$ and $Li_2(VO)(Si,Ge)O_4$. However, there is no competition between $J_2$ and $J_4$ interactions in a lattice of smaller squares, since both these interactions are ferromagnetic. In a recently published paper [22], the compound $Zn_2(VO)(PO_4)_2$ was investigated, and it was stated that $J_2$ and $J_4$ interactions are, on the contrary, antiferromagnetic, and diagonal $J_4$ interactions are very weak ($J_4/J_2$ = 0.02).

Unlike these compounds, the absolute value of frustration ratio α ($J_4/J_2$) in $(VO)SiP_2O_8$ and $(VO)(H_2PO_4)_2$ is less than the critical value ($\alpha = 1/2$) and is equal to 0.3. One should mention that in $(VO)SiP_2O_8$ both interactions $J_2$ and $J_4$ are AF and compete with one another, whereas in $(VO)(H_2PO_4)_2$ there is no competition between these interactions, since $J_4$ is FM while $J_2$ is AF.

As was shown above, in all the compounds, except $Zn_2(VO)(PO_4)_2$, the lattices of smaller squares formed by $J_2$ interactions are to some extent distorted and are not square. The regular square lattices are formed by the nearest-neighbor $J_4$ (parameters of unit cell a and b) and by the second neighbor (diagonal) $J_8$ interactions in the ab plane (figure 1(a), (c), (e)). The $J_8$ interactions are AF in all compounds; hence, these lattices of larger squares would be frustrated irrespective of the sign of $J_4$ interactions in the case of tetragonal symmetry of crystals. The absolute value of the α ratio ($J_8/J_4$) is less than critical value in $(VO)SO_4$ (0.1) and $Li_2(VO)GeO_4$ (0.4) while it is approximately equal to 1/2 in $Li_2(VO)SiO_4$ and exceeds critical value in $Zn_2(VO)(PO_4)_2$ (0.9), $(VO)SiP_2O_8$ (1.3), $(VO)MoO_4$ (1.7), and $(VO)(H_2PO_4)_2$ (4.6).

Thus, virtually all the magnetic interactions in these compounds compete with each other. Determination of the magnetic state of these systems is a very complicated problem, since more than one specific geometrical configuration of $V^{4+}$ ions causes competition of magnetic interactions.

## 4. Conclusions

The crystal chemical method enabled us to use the structural data to calculate the sign and strength of magnetic interactions of not only of the nearest neighbors, but also of the next-nearest neighbors in the tetragonal compounds $Zn_2(VO)(PO_4)_2$, $(VO)(H_2PO_4)_2$, $(VO)SiP_2O_8$, $(VO)SO_4$, $(VO)MoO_4$, $Li_2(VO)SiO_4$ and $Li_2(VO)GeO_4$ with similar sublattice of $V^{4+}$ ions and to establish the origin of the differences between the compounds. The distinctive feature of these compounds is a strong dependence of the characteristics of magnetic interactions on slight displacement of $XO_4$ (X = P, Mo, Si or Ge) groups. This is related to the fact that the ions of the same $XO_4$ group are intermediate ions in several interactions simultaneously and quite often they are in the space of interaction between the magnetic ions in critical positions, and any slight deviations from these positions can result in a change of the sign or a drastic change of the strength of magnetic interaction. This dependence is intensified in addition by the "rigidity" of X–O bonds in $XO_4$ groups. Therefore, the effect of substitution, temperature or pressure can result in displacement or rotation of the group as a whole unit.

Another characteristic of these compounds is related the fact that sublattices of magnetic $V^{4+}$ ions in them consist of a great number of specific geometrical configurations, in which a competition of magnetic interactions can occur. However, in the literature the competition of magnetic interactions is mainly considered only in one fragment – a lattice of smaller distorted squares. Although ambiguous data concerning the frustration ratio $\alpha$ ($J_4/J_2\{J_2/J_1\}$) between the nearest-neighbor coupling $J_2\{J_1\}$ and the second-neighbor (diagonal) coupling $J_4\{J_2\}$ in the square lattices were obtained for $(VO)MoO_4$, $Li_2(VO)SiO_4$ and $Li_2(VO)GeO_4$, nonetheless, as small ($1 \leq \alpha \leq 4$) as large ($5 \leq \alpha \leq 12$) values of $\alpha$ allow interpreting the experimental data [1, 2, 5]. We have shown that the reason for this ambiguous character is concerned with the fact that interactions between the same $V^{4+}$ ions form another, more distorted, square lattice perpendicular to the *c*-axis, with the same diagonal coupling $J_4\{J_2\}$, but with another nearest-neighbor coupling $J_3$ (figure 1). Here, according to our calculations, for the first (less distorted) lattice the values $\alpha$ are small while for the second one (more distorted) they are large.

The analysis of the structure of $V^{4+}$ sublattices in these compounds has shown that, beside the widely researched lattice of smaller distorted squares, there are also other specific geometrical configurations of ions $V^{4+}$, in which characteristic properties of magnetic interactions cannot be exhibited simultaneously. Virtually all magnetic interactions in these compounds compete among themselves, including the AF nearest- and next-nearest-neighbor interactions ($J_7^s / J_1^s$ ~ 1/6) in the linear VO–VO chains along the *c*-axis. To understand the magnetic state of systems studied it is not sufficient to consider the competition of magnetic interactions only on one configuration; it is also necessary to take into account the mutual influence of competition of magnetic interactions on various specific geometrical configurations of $V^{4+}$ ions.

**Acknowledgment**

This work is supported by grant 06-I-P8-009 of Far Eastern Branch of Russian Academy of Sciences.